\documentclass[aps,prl,twocolumn,groupedaddress,showpacs]{revtex4}

 \usepackage{graphicx}
 \usepackage{dcolumn}
 \usepackage{bm}
 \usepackage{color}

\begin{document}


\title{Bi(111) thin film with insulating interior but metallic surfaces}

\author{Shunhao Xiao}%
\author{Dahai Wei}%
\author{Xiaofeng Jin}%

\email[Corresponding author, E-mail: ]{ xfjin@fudan.edu.cn }
\affiliation{ State Key Laboratory of Surface Physics and Department of Physics, Fudan University, Shanghai
200433, China }

\date{\today}

\begin{abstract}

The electrical conductance of molecular beam epitaxial Bi on BaF$_{2}$(111) was measured as a function of both film thickness (4-540 nm) and temperature (5-300 K). Unlike bulk Bi as a prototype semimetal, the Bi thin films up to 90~nm are found to be insulating in the interiors but metallic on the surfaces. This result has not only resolved unambiguously the long controversy about the existence of semimetal-semiconductor transition in Bi thin film but also provided a straightforward interpretation for the long-puzzled temperature dependence of the resistivity of Bi thin films, which in turn might suggest some potential applications in spintronics.

\end{abstract}

\pacs{75.30.Et, 75.30.Gw}

\maketitle




Bi is a pentavalent element in the periodic table with the atomic structure of [Xe]4f$^{14}$5d$^{10}$6s$^{2}$6p$^{3}$. It crystallizes in a rhombohedral (A15) structure with two ions and ten valence electrons per primitive cell. The even number of valence electrons makes it very close to being insulator, but the very slight overlap between the conduction and valence bands eventually drives it to a prototype semimetal with a very small number of carriers ( 3$\times$10$^{17}$~cm$^{-3}$), therefore leading to an unusually long Fermi wavelength (30~nm) \cite{refmermin, refPhHofmannReview}.

It was predicted theoretically that due to the quantum size effect, Bi should undergo a semimetal to semiconductor transition in thin films when the thicknesses are comparable to the Fermi wavelength \cite{SMSCtheorylutskii,SMSCtheorySandomirskii}. However, the experimental identification is highly nontrivial and remains contradictory, although considerable effort has been made during the last fifty years \cite{duggal:492,refsmsc,refcomSMSC,refReplyComSMSC,refPhHofmannReview}. Especially, the experimentally observed non-monotonic temperature dependence of the electrical resistivity does not fit at all to the physical picture anticipated by the existence of a semiconductor phase \cite{refR.A.HoffmanBimica, refNGarcia, refCLQianScience, refBiBaF2}. It was argued recently by Hirahara et al., by means of angle resolved photoemission (ARPES), that Bi(111) films should always be metallic because of the persistence of thickness-independent metallic surface states, in direct contrast to the semimetal-semiconductor transition \cite{refBiSiARPES, refSiSS}. Therefore,  the existence of semimetal-semiconductor transition in Bi films remains as a puzzle.

In this letter, we are going to demonstrate unambiguously that the interior of a Bi(111) film thinner than 90 nm prepared by molecular beam epitaxy on BaF$_{2}$(111) is indeed insulating or semiconducting while its surfaces are metallic. This result has finally clarified the long-standing issue of the semimetal-semiconductor transition in Bi films; in addition it might be related and even provide a new playground to explore the topological property of Bi \cite{refTIBimurakami,refTIBi,WuYongshi} and Bi-based topological insulators \cite{FuLiangKane,HasanBiSbNature,refTI}.

Single crystalline Bi(111) films ranging from 4 to 540 nm thick were epitaxially
grown on BaF$_{2}$(111) by molecular-beam epitaxy in an ultra high vacuum system equipped with reflection
high energy electron diffraction (RHEED) and Auger electron spectroscopy (AES) \cite{PhysRevLett.94.137210}. Clean and ordered BaF$_{2}$(111) substrates were first prepared by annealing at $700~\,^{\circ}\mathrm{C}$ for 10 minutes then cooled down and kept at $70~\,^{\circ}\mathrm{C}$, on which the epitaxial growth of Bi was followed while the evaporation rate was monitored by a quartz microbalance. To prevent oxidation in ambient air during the transport measurement, a 6~nm MgO capping layer was deposited on each sample before taken out from the UHV chamber. The grazing angle X-ray diffraction (XRD) experiment was performed at Beijing Synchrotron Radiation Facility \cite{PhysRevLett.97.067203}. The the transport measurements were carried out on the samples  patterned into standard Hall bars along the [11$\bar{2}$] direction, using Quantum Design physical property measurement system (PPMS-9T) \cite{PhysRevLett.103.087206}.

\begin{figure}
\includegraphics[width=9cm]{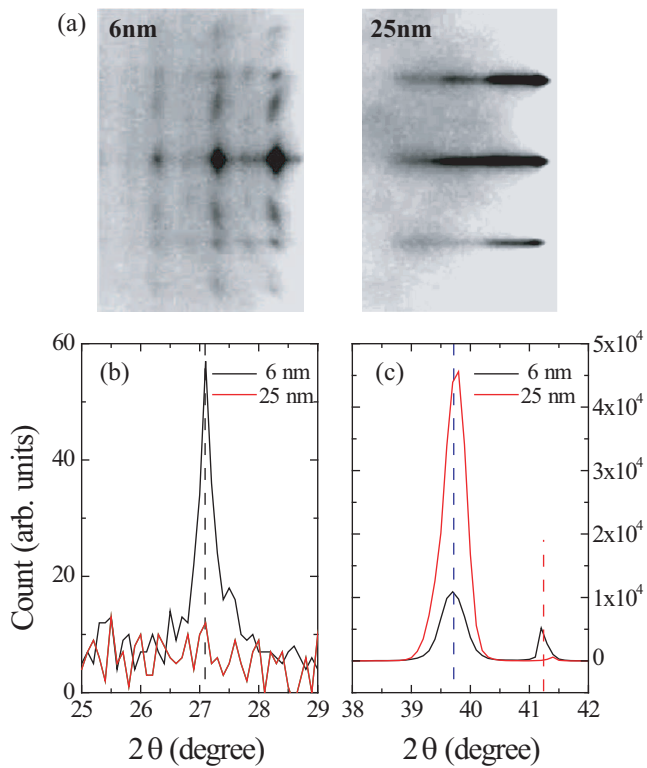}
\caption{\label{fig1}(a) RHEED patterns of 6 and 25~nm of Bi on BaF$_{2}$; the incident electron beam is along the $[11\bar{2}]$ direction of BaF$_{2}$. (b) The grazing angle XRD spectra of the same samples; the small peak around 41~$^{\circ}$ comes from the (0$\bar{2}$2) diffraction of the BaF$_{2}$ substrate.
}
\end{figure}

Fig.~\ref{fig1} shows a set of representative RHEED and grazing angle XRD results for Bi/BaF$_{2}$(111) at different film thicknesses, which clearly indicates that the epitaxial growth of Bi on BaF$_{2}$(111) involves two stages. In the early stage below about 15 nm, the RHEED patterns (as for 6 nm thick film) are complicated, implying that the Bi films grow in polycrystal. This is confirmed by the XRD result which shows the coexistence of Bi pseudo-cubic phase ((100) peak at 27.1 degree) and Bi hexagonal phase ((1100) peak at 39.7 degree) in Fig.~\ref{fig1}(b) and (c), respectively. In the later stage of film growth beyond 15~nm, the RHEED patterns (as for 25 nm thick film) become very simple, implying that the Bi films grow in single crystal. This is again revealed by XRD that the pseudocubic phase has disappeared and transformed completely to the hexagonal phase of Bi(111) as seen in Fig.~\ref{fig1}(b) and (c). In fact, the overall growth behavior realized here in Bi/BaF$_{2}$(111) is quite similar to the previous observation of Bi on Si(111)~7$\times$7 \cite{refBiSi}, although the structure transformation there happened at thinner film thickness. We there conclude that we can obtained experimentally pure Bi(111) single crystal films thicker than 15 nm on BaF$_{2}$(111).

\begin{figure}
\includegraphics[width=8cm]{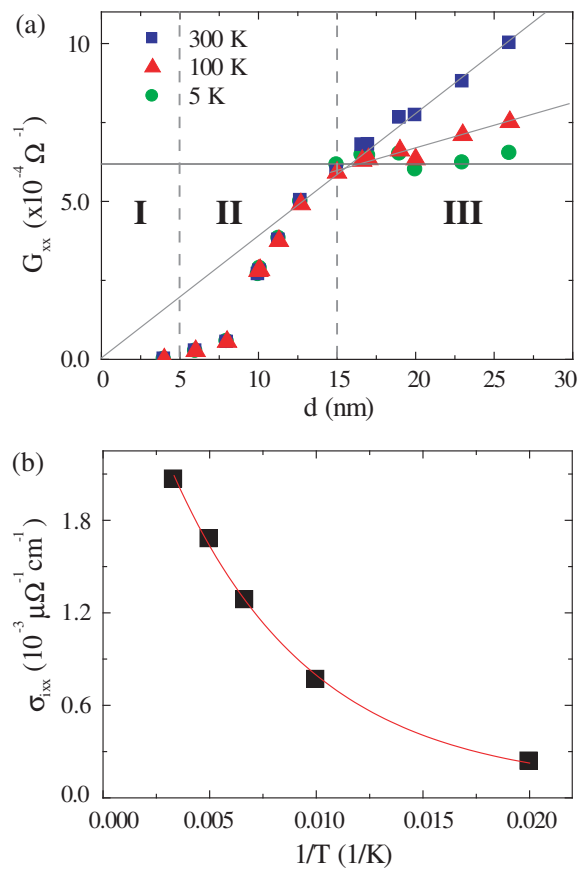}
\caption{\label{fig2}(a) Conductance as a function of the sample thickness at 300~K, 100~K and 5~K.
(b) Conductivity of the film interior derived from the slops of the conductance-thickness lines in region (III) of (a) as a function of temperature. The curve is an exponential fitting.}
\end{figure}

Fig.~\ref{fig2}(a) shows the electrical conductance  of Bi as a function of film thickness measured at different temperatures, where three distinct regimes as marked can be realized. In regime (I) the conductance is extremely small reflecting the insulating nature of the pesudocubic phase of Bi \cite{refSiSS}. Regime (II) corresponds to the mixture of both pesudocubic and hexagonal phases of Bi, and the transformation from the former to the latter. The most interesting and surprising result lies in regime (III), where the conductance exhibits strong temperature dependence. It is noted that at 5 K the conductance remains almost unchanged with increasing Bi film thickness from 15 nm to 25 nm, i.e., the additional 10 nm Bi doesn't contribute at all to the electrical conductance; this in turn strongly implies that the low temperature electrical transport in Bi films should be dominated by the metallic surface state while the corresponding film interiors are insulating. In contrast to the 5~K case the conductance at 300 K increases linearly as a function of film thickness, which suggests that besides the surface channel the electron transport in the film interior has also contributed to the conductance. However, this is possible only when the film interior is an insulator or a semiconductor with a very small energy gap about 10 meV, then it would behave like an insulator at low temperature (e.g., 5 K) because of the energy gap, but would become conducting at higher temperatures (e.g., 300 K) when the thermal excitation from the valence band to conduction band is no longer negligible. Therefore, this is the first direct experimental evidence for the existence of the long debated insulating or semiconducting phase in Bi films, although the physical picture turns out to be more complicated than what has been originally proposed. It is clearly now that Bi(111) thin films with thicknesses comparable the Fermi wavelength are indeed insulating or semiconducting but only in the film interiors yet their surfaces are always metallic.

Now we turn to estimate quantitatively the energy band gap of the Bi(111) film interior. Because of the two channels for the total conductance: the surface and interior contributions, we have:
\begin{equation}\label{G}
    G_{xx}=G_{s}+\sigma_{ixx}\frac{w}{l}d
\end{equation}

$G_{xx}$ and $G_{s}$ are the total and surface conductances respectively, and $\sigma_{ixx}$ is the conductivity of the film interior; $d$, $l$, $w$ are the thickness, length, and width of the Bi(111) Hall bar respectively. It is immediately recognized from Eq.~\ref{G} that $G_{xx}$ is proportional to $d$ at any given temperature, assuming $\sigma_{ixx}$ is a constant in the thickness range of 15 to 25 nm, which explains nicely the observations in Fig.~\ref{fig2}(a). By getting the slops at different temperatures, we establish a relation between $\sigma_{ixx}$ and temperature, as shown in Fig.~\ref{fig2}(b) as  $\sigma_{ixx}$ $vs$ $1/T$. This set of data can be well fitted by an exponential decay function $\alpha\cdot e^{-\frac{\Delta E}{kT}}$; here $\alpha$ is a constant and $\Delta E$ is the energy gap to be determined, while $k$ and $T$ are the Boltzmann constant and temperature respectively. The nice fit of the experimental data with this gapped exponential function leads to an energy gap of $\Delta E$=12~meV. It is such a small energy gap that results in the insulating behavior of the Bi(111) film interior at 5 K but the conducting behavior at higher temperatures.

\begin{figure}
\includegraphics[width=8cm]{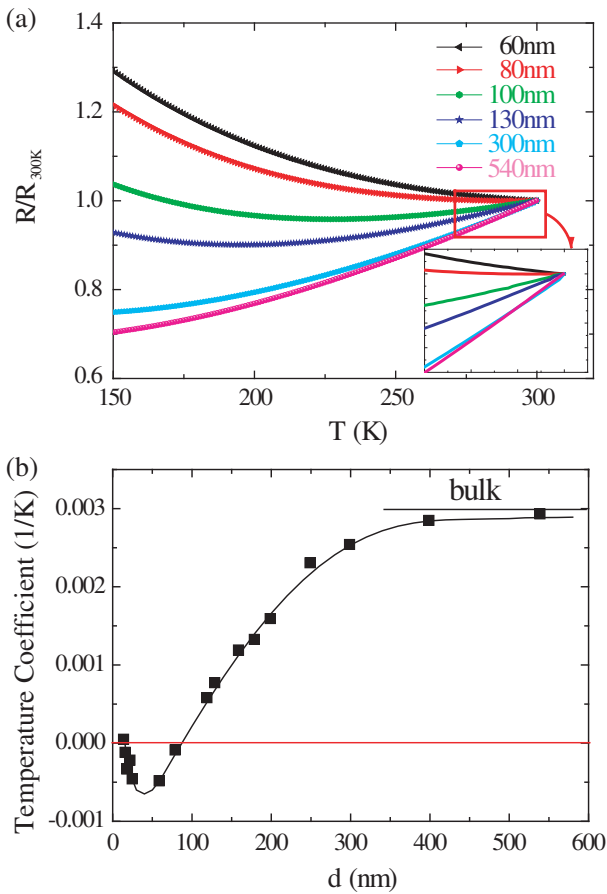}
\caption{\label{fig3}(a) Resistance-temperature curves normalized with the value at 300~K. The inset is a zoom-in from 270~K to 300~K.
(b) The temperature coefficients of resistivity between 270~K and 300~K as a function of sample thicknesses. The black curve is a guide to the eye.}
\end{figure}

After identifying the semiconductor phase in single crystalline Bi films, next we try to find out how thin of a Bi(111) film that would undergo the semimetal-semiconductor transition. In Fig.~\ref{fig3}(a), the resistance $vs$ temperature curves normalized with the value at 300~K were shown. It is easily seen from the inset that the slope of the curves changes sign from positive to negative as the temperature is decreased - an unambiguous evidence for the semimetal to semiconductor transition in Bi thin films. By plotting the temperature coefficient of resistivity around 300 K ($\frac{\rho_{300K}-\rho_{270K}}{\rho_{300K}\cdot30~K}$) as a function of film thickness in Fig.~\ref{fig3}(b), we find that the temperature coefficient crosses over zero at about 90 nm, which defines exactly the semimetal to semiconductor transition in Bi(111) films. This result agrees well with an earlier observation in ref.~\cite{duggal:492}. Presumably due to the poor sample quality especially the surface quality, what has missed in that work is the well-defined dip appearing at thinner Bi film thicknesses as clearly seen in Fig.~\ref{fig3}(b) in our single crystalline films by molecular beam epitaxy. The appearance of the dip is in fact a result of the competition between the metallic surface and the semiconductor interior of Bi(111) thin film according to Eq.~\ref{G}, as the relative weight of the former gradually increases and eventually dominates as the film thickness decreases.

\begin{figure}
\includegraphics[width=8.5cm]{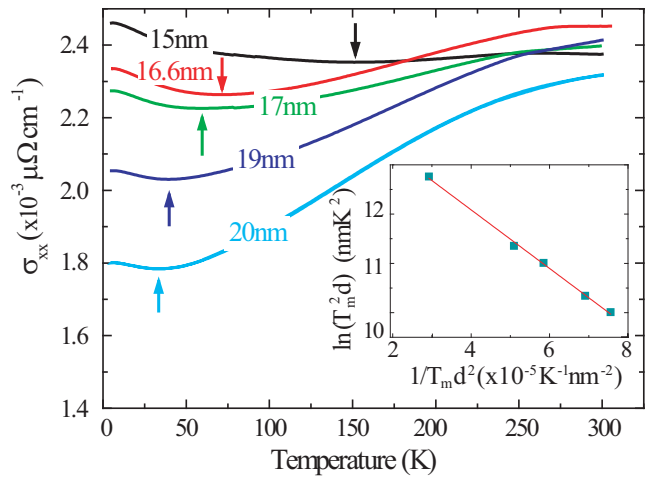}
\caption{\label{fig4}Temperature dependent conductivities of samples with different thickness. The inset shows ln(T$_{m}^{2}$d) as a function of 1/T$_{m}$d$^{2}$.
}
\end{figure}

With the new understanding on the Bi(111) thin film discussed above, we are going to demonstrate in the following that the longstanding puzzle about the non-monotonic behavior of the resistivity or conductivity versus temperature curve in Bi films can also be resolved. Similar to Eq.~\ref{G}, the total conductivity $\sigma_{xx}$ from the two (surface and film interior) channels can be expressed as
\begin{equation}\label{equ2}
    \sigma_{xx}=\frac{\sigma_{s}}{d}+\alpha\cdot e^{-\frac{\Delta E}{kT}}
\end{equation}
Here $\sigma_{s}$ is the surface conductivity. According to the Matthiessen's rule, it can be further expressed in resistivity as $\sigma_{s}=\frac{1}{(\rho_{0}+\rho_{T})}$, where $\rho_{0}$ and $\rho_{T}$ are the surface residual and electron-phonon induced resistivity. By adopting the results in literature for the surface electron-phonon induced resistivity $\rho_{T}=sT$ \cite{PhysRevLett.99.146805}, and the quantum size effect induced energy gap $\Delta E=\frac{b}{d^{2}}$ ($b$ is a constant) \cite{SMSCtheorySandomirskii,SMSCtheorylutskii}, then Eq.~\ref{equ2} turns into
\begin{equation}\label{equ3}
    \sigma_{xx}=\frac{1}{d(\rho_{0}+sT)}+\alpha\cdot e^{-\frac{b}{kTd^{2}}}
\end{equation}

 For a given sample with fixed film thickness, it is obvious that the first term from the metallic surface would decrease as the sample temperature increases, but the second term from the semiconductor interior would increase. It is exactly this competition that caused the nonmonotonic behavior of the temperature dependent conductivity or resistivity. Depending on Bi film thickness, when the two terms are comparable in magnitude, a valley in conductivity should be expected in principle, which would qualitatively explain the experimentally observed phenomena in Fig.~\ref{fig4}; this complicated behavior of conductivity versus temperature as well as film thickness has always made the electrical transport of Bi films puzzled and mysterious. Quantitatively, for each fixed film thickness the valley as a function of temperature can be determined by minimizing Eq.~\ref{equ3} while noticing the experimental fact that $\rho_{T}$ is smaller than $\rho_{0}$, therefore we reach to the following equation:

\begin{equation}\label{equ4}
    \ln(T_{m}^{2}d)=-\frac{b}{kT_{m}d^{2}}+\ln(\frac{\alpha b \rho_{0}^{2}}{sk})
\end{equation}
Clearly, as seen in the inset of Fig.~\ref{fig4}, the experimental data of $\ln(T_{m}^{2}d)$ $vs$ $\frac{1}{T_{m}d^{2}}$ can be well fitted by a straight line with a slop of $-b/k\approx-6\times10^4$~K$\cdot$nm$^2$, by which we can get the film thickness dependent energy gaps of the semiconducting interiors:  23~meV for 15~nm and 13~meV for 20~nm Bi films, respectively, which not only agree very well with the foregoing obtained result deduced from Fig.~\ref{fig2}(b) but also provide the trend of thickness dependent
energy gap of the film interiors in Bi.

In summary, we have demonstrated in this work that the long debated semimetal to semiconductor transition does happen in Bi(111) thin film when the film thickness is comparable to the Fermi wavelength of Bi; the problem had been controversial because of the subtle fact that the film interior is semiconducting while the surface is always metallic.

\begin{acknowledgments}
This work was supported by MSTC (No. 2009CB929203 and No. 2006CB921303), NSFC
(No. 10834001), and SCST. The technical support from Beijing Synchrotron Radiation Facility (BSRF) for XRD measurements is also acknowledged.
\end{acknowledgments}


\end{document}